\documentclass[12pt]{iopart}

\usepackage{iopams,graphicx,amssymb}
\eqnobysec

\begin{document}

\title[Anomalous scaling of passively advected vector fields] {Anomalous
scaling in statistical models of passively advected vector fields}

\author{N. V. Antonov and N. M. Gulitskiy}

\address{Department of Theoretical Physics, St.~Petersburg University,
Uljanovskaja 1, St.~Petersburg, Petrodvorez, 198504 Russia}

\ead{nikolai.antonov@pobox.spbu.ru, ngulitskiy@gmail.com}

\begin{abstract}
The field theoretic renormalization group and the operator product expansion
are applied to the stochastic model of passively advected vector field with
the most general form of the nonlinear term allowed by the Galilean
symmetry. The advecting turbulent velocity field is governed by the
stochastic Navier--Stokes equation. It is shown that the correlation
functions of the passive vector field in the inertial range exhibit
anomalous scaling behaviour. The corresponding anomalous exponents are
determined by the critical dimensions of tensor composite fields (operators)
built solely of the passive vector field. They are calculated
(including the anisotropic sectors) in the leading order of the expansion
in $y$, the exponent entering the correlator of the stirring force in the
Navier--Stokes equation (one-loop approximation of the renormalization
group). The anomalous exponents exhibit an hierarchy related to the
degree of anisotropy: the less is the rank of the tensor operator, the
less is its dimension. Thus the leading terms, determined by scalar
operators, are the same as in the isotropic case, in agreement
with the Kolmogorov's hypothesis of the local isotropy restoration.
\end{abstract}

\section{Introduction} \label{sec:Intro}

In the past two decades, much attention has been attracted by turbulent
advection of passive scalar fields; see the review paper \cite{FGV} and
references therein. Being of practical importance in itself, the problem
of passive advection can be viewed as a starting point for studying
intermittency and anomalous scaling in the fluid turbulence on the whole
\cite{Legacy}. Most progress was achieved for the so-called Kraichnan's
rapid-change model: for the first time, the anomalous exponents were
derived on the basis of a microscopic model and within controlled
approximations \cite{GK}. In Kraichnan's model the advecting velocity field
$v_{i}(x)$ with $x = \{t,{\bf x}\}$ is modelled by a Gaussian statistics
with vanishing correlation time and prescribed correlation function
$\langle vv\rangle \propto \delta(t-t') k^{-d-\xi}$, where $k$ is the wave
number, $d$ is the dimension of space and $\xi$ is an arbitrary exponent
with the most realistic (Kolmogorov) value $\xi=4/3$.

The ``zero-mode approach,'' developed in \cite{GK}, is based on the fact
that, due to the temporal decorrelation of the Kraichnan's velocity field,
some closed differential equations can be derived for the equal-time
correlation functions of the passive fields. In this sense, the model
is equivalent to a certain quantum-mechanical problem and appears
``exactly solvable.'' Although such equations cannot be explicitly solved,
the anomalous exponents can be extracted from the analysis of their
infrared (IR) asymptotic behaviour; see~\cite{FGV} for a detailed review
and the references.

In \cite{RG} and subsequent papers, the field theoretic renormalization
group (RG) and the operator product expansion (OPE) were applied to
Kraichnan's model; see \cite{JPhysA} for a review and the references.
In that approach, the anomalous scaling emerges as a consequence of the
existence in the corresponding OPE of certain composite fields (``operators''
in the quantum-field terminology) with {\it negative} dimensions, which are
identified with the anomalous exponents. This allows one to
construct a systematic perturbation expansion for the anomalous exponents
and to calculate them up to the order $\xi^{2}$ \cite{RG} and $\xi^{3}$
\cite{cube}. Besides the calculational efficiency, an important advantage
of the RG+OPE approach is its relative universality: it can also be applied
to the case of finite correlation time or non-Gaussian advecting fields.

For passively advected {\it vector} fields, any calculation of the exponents
for higher-order correlations calls for the RG techniques already in the
$O(\xi)$ approximation.

Owing to the presence of a new stretching term in the dynamic equation, the
 behavior of the passive  vector field appears much richer than that of
the scalar field: ``...there is considerably more life in the large-scale
transport of vector quantities'' (p. 232 of Ref. \cite{Legacy}). Indeed,
passive vector fields reveal anomalous scaling already on the level of the
pair correlation function \cite{MHD,Lanotte,MHD1}. They also exhibit
interesting
large-scale instabilities that can be interpreted as manifestation of the
dynamo effect \cite{MHD,Ka68,DV}. Some special case (${\cal A}=0$, see
below) reveals a close formal resemblance to the NS turbulence
\cite{matrix,rebumat,Sh}. From the physics viewpoints, passive
vector fields can have different physical meaning: magnetic field,
weak perturbation of the prescribed background flow, concentration
or density of the impurity particles with an internal structure.

In this paper, we study anomalous scaling of a passive vector quantity,
advected by a non-Gaussian velocity field with finite correlation time,
governed by the stirred Navier--Stokes (NS) equation.

The plan of the paper is the following.
In sec.~\ref{sec:Model} we give detailed description of the stochastic
problem and explain the motivation of our study and its relation to
previous work. In sec.~\ref{sec:QFT} we give the field theoretic formulation
of the model. In sec.~\ref{sec:RG} we establish its renormalizability,
derive the corresponding RG equations and present the explicit one-loop
results for the renormalization constants and RG functions.
Possible IR attractive fixed points are discussed in sec.~\ref{sec:FP}.
In sec.~\ref{sec:OPE} the operator
product expansion is employed to establish the anomalous scaling of the
correlation functions in the inertial-range. The corresponding anomalous
exponents are determined by the critical dimensions of tensor composite
operators built solely of the passive field. The practical calculation
is performed in the leading (one-loop) approximation; the results are
presented in sec.~\ref{sec:V}. Section ~\ref{sec:Conc} is reserved for a
brief conclusion; in particular, we mention an hierarchy demonstrated
by the anisotropic contributions.

\section{Description of the model} \label{sec:Model}

We confine ourselves with the case of transverse (divergence-free) passive
$\theta_{i} (x)$ and advecting $v_{i}(x)$ vector fields.
Then the general advection-diffusion equation has the form
\begin{eqnarray}
\nabla _t\theta_{i} - {\cal A}_{0} (\theta_{k}\partial_{k}) v_{i} +
\partial_{i} {\cal P} = \kappa_0\partial^{2} \theta_{i} + \eta_{i},
\qquad
\nabla_t \equiv \partial _t + (v_{k}\partial_{k}) ,
\label{1}
\end{eqnarray}
where $\nabla_t$ is the Lagrangian (Galilean covariant) derivative,
${\cal P}(x)$ is the pressure, $\kappa_0$ is the diffusivity,
$\partial^{2}$ is the Laplace operator and $\eta_{i}(x)$ is a transverse
Gaussian stirring force with zero mean and covariance
\begin{equation}
\langle \eta_{i}(x) \eta_{k}(x')\rangle = \delta(t-t')\, C_{ik}(r/L).
\label{2}
\end{equation}
The parameter $L$ is an integral scale related to the stirring, and $C_{ik}$
is a dimensionless function with the condition $\partial_{i}C_{ik}=0$, finite
at $r=0$ and rapidly decaying for $r\to\infty$; its precise form is
unessential. Due to the transversality conditions $\partial_{i}\theta_{i}=0$,
$\partial_{i}v_{i}=0$, the pressure can be expressed as the solution of the
Poisson equation:
\begin{equation}
\partial^{2} {\cal P} = ({\cal A}_{0} - 1) \,
\partial_{i} v_{k}  \partial_{k} \theta_{i}.
\label{Pois}
\end{equation}
Thus the pressure term makes the dynamics (\ref{1}) consistent with the
transversality. The amplitude factor ${\cal A}_{0}$ in front of the
``stretching term'' $(\theta_{k}\partial_{k}) v_{i}$ is not fixed by the
Galilean symmetry and thus can be arbitrary. Such general ``${\cal A}$
model'' was introduced and studied in refs. \cite{amodel}--\cite{HA};
it can be naturally justified within the so-called multiscale techniques.

From the physics viewpoints most interesting is the special case
${\cal A}_{0}=1$, where the pressure term disappears: it corresponds to
magnetohydrodynamic (MHD) turbulence \cite{MGD}. It was studied earlier in
numerous papers; see e.g. refs. \cite{MHD,Lanotte,MHD1,AG,JJ}
and references therein.

In earlier studies, the velocity field in (\ref{1}) was usually described
by the Kraichnan's rapid-change model: Gaussian statistics with vanishing
correlation time and prescribed power-like correlation function.
In this paper, we employ the stochastic
NS equation:
\begin{equation}
\nabla _t v_i=\nu _0\partial^{2} v _i-\partial _i {\wp}+f_i,
\label{1.1}
\end{equation}
where $\nabla _t$ is the same Lagrangian derivative, ${\wp}$ and $f_i$
are the pressure and the transverse random force per unit mass. We assume
for $f$ a Gaussian distribution with zero mean and correlation function
\begin{equation}
\big\langle f_i(x)f_j(x')\big\rangle = \frac{\delta (t-t')}{(2\pi)^{d}}\,
\int_{k\ge m} d{\bf k}\, P_{ij}({\bf k})\, d_f(k)\, \exp \big[{\rm i}{\bf k}
\left({\bf x}-{\bf x}'\right)\big] ,
\label{1.2}
\end{equation}
where $P_{ij}({\bf k}) =\delta _{ij}  - k_i k_j / k^2$ is the transverse
projector, $d_f(k)$ is some function of $k\equiv |{\bf k}|$ and model
parameters. The momentum $m=1/L$, the reciprocal of the integral scale
$L$ related to the velocity, provides IR regularization. For simplicity,
we do not distinguish it from the integral scale related to the scalar
noise in (\ref{2}).

The standard RG formalism is applicable to the problem (\ref{1.1}),
(\ref{1.2}) if the correlation function of the random force is chosen
in the power form \cite{DeDom}
\begin{equation}
d_f(k)=D_0\,k^{4-d-y},
\label{1.9}
\end{equation}
where $D_{0}>0$ is the positive amplitude factor and the exponent
$0<y\le 4$ plays the role of the RG expansion parameter. The
most realistic value of the exponent is $y=4$: with an
appropriate choice of the amplitude, the function (\ref{1.9}) for
$y\to4$ turns to the delta function, $d_f(k) \propto
\delta({\bf k})$, which corresponds to the injection of energy to
the system owing to interaction with the largest turbulent eddies;
for a more detailed justification see e.g. \cite{Book3,turbo}.

\section{Field theoretic formulation} \label{sec:QFT}

According to the general theorem (see e.g. \cite{Book3}), the full-scale
stochastic problem (\ref{1})--(\ref{1.9}) is equivalent
to the field theoretic model of the doubled set of fields
$\Phi=\{v,v',\theta,\theta'\}$ with the action functional
\begin{equation}
{\cal S} (\Phi )= {\cal S}_{v}({\bf v}', {\bf v}) +
\theta' D_{\theta} \theta'/2 +
\theta' \left\{ -\nabla_{t} - {\cal A}_{0} (\theta_{k}\partial_{k}) v_{i}
+ \kappa_{0} \partial^{2} \right\} \theta,
\label{action}
\end{equation}
where $D_{\theta}$ is the correlation function (\ref{2}) of the random
force $\eta_{i}$ in (\ref{1}) and $S_{v}$ is the action for the problem
(\ref{1.1})--(\ref{1.9}):
\begin{equation}
{\cal S}_{v}({\bf v}', {\bf v}) =
v 'D_{v}v'/2+ v'\left\{-\nabla_t + \nu_0 \partial^{2} \right\}v ,
\label{actionV}
\end{equation}
where $D_{v}$ is the correlation function (\ref{1.2}) of the random force
$f_{i}$. All the integrations over $x=\{t,{\bf x}\}$ and summations
over the vector indices are understood. The auxiliary vector fields
$v',\theta'$ are also transverse,
$\partial_{i}v_{i}'=\partial_{i}\theta_{i}'=0$, which allows to omit the
pressure terms on the right-hand sides of expressions (\ref{action}),
(\ref{actionV}), as becomes evident after the integration by parts.
For example,
\[ \int dt \int d{\bf x} \ v_{i}'\partial_{i} {\wp} = - \int dt
\int d{\bf x} \ {\wp} (\partial_{i}v_{i}') =0 . \]
Of course, this does not mean that the pressure contributions are
unimportant: the fields $v',\theta'$ act as transverse projectors and
select the transverse parts of the expressions to which they are contracted.

The part of the coupling constants is played by the three parameters
$g_{0}\equiv D_{0}/\nu_0^3$, ${\cal A}_{0}$
and $u_{0} = \kappa_{0}/\nu_0$, the analog of
the inverse Prandtl number in the scalar case. By dimension,
\begin{equation}
g_{0} \propto \Lambda^{y}, \quad
{\cal A}_{0}\ {\rm and}\  u_{0} \propto \Lambda^{0},
\label{Lambda}
\end{equation}
where $\Lambda$ is the characteristic ultraviolet (UV) momentum scale. Thus
the model (\ref{action}), (\ref{actionV}) becomes logarithmic (all the
coupling constants become dimensionless) at $y=0$, and the UV divergences
manifest themselves as poles in $y$.

\section{Renormalization and RG equations} \label{sec:RG}

The renormalization and RG analysis of the model (\ref{action}),
(\ref{actionV}) are similar to that of the scalar advection by the NS
velocity field \cite{NSpass}, and here we discuss them only briefly.
Dimensional analysis shows that superficial UV divergences can be
present only in the 1-irreducible Green functions $\langle v'v \rangle$,
$\langle v'vv \rangle$, $\langle\theta'\theta\rangle$ and
$\langle\theta'v\theta\rangle$.
The corresponding counterterms reduce to the forms
$v'\partial_{t}v$,  $v'\partial^{2}v$, $v'(v\partial)v$,
$\theta'\partial_{t}\theta$, $\theta'\partial^{2}\theta$,
$\theta' (\theta\partial) v$ and $\theta' (v\partial) \theta$.

The spatial derivative $\partial$ at the vertices $v'(v\partial)v$,
$\theta' (\theta\partial) v$ and $\theta' (v\partial) \theta$
in (\ref{action}) can be moved, using the integration by parts, onto
the auxiliary fields $v'$ and $\theta'$. Thus any counterterm must include
one spatial derivative per each auxiliary field. This excludes the
counterterms $v'\partial_{t}v$ and $\theta'\partial_{t}\theta$ without
a spatial derivative. Then the Galilean symmetry excludes the structures
$v'(v\partial)v$ and $\theta' (v\partial) \theta$ because they must enter
the counterterms only in the form of Galilean invariant combinations
$v'\nabla_{t}v$ and $\theta'\nabla_{t}\theta$.

The remaining three counterterms $v'\partial^{2}v$,
$\theta'\partial^{2}\theta$ and $\theta' (\theta\partial) v$
can be reproduced by multiplicative renormalization of the parameters
\begin{equation}
\nu_0 = \nu Z_{\nu}, \quad \kappa_0 = \kappa Z_{\kappa}, \quad
{\cal A}_{0} = {\cal A} Z_{\cal A}, \quad
g_{0} = g \mu^{y} Z_{g}, \quad Z_{g} =  Z_{\nu}^{-3};
\label{mult}
\end{equation}
no renormalization of the fields $\Phi=\{v,v',\theta,\theta'\}$ and the IR
scale $m$ is needed. Here $\nu$, $g$, $\kappa$ and ${\cal A}$ are
renormalized analogs of the bare
parameters $\nu_{0}$, $g_{0}$, $\kappa_{0}$ and ${\cal A}_{0}$, while
the reference scale $\mu$ is an additional parameter of the renormalized
theory. The last relation in (\ref{mult}) follows from the absence of
renormalization of the amplitude $D_{0}=g_{0}\nu_{0}^{3}= g\mu^y \nu^3$ in
the first term of the action ${\cal S}_{vR}$. The renormalization constants
$Z_{i}=Z_{i}(g,u,{\cal A},d,y)$ absorb all the UV divergences, so that the
Green functions are UV finite (that is, finite at $y=0$) when expressed
in renormalized parameters.

The corresponding renormalized action has the form
\begin{eqnarray}
{\cal S}_{R} (\Phi )= {\cal S}_{vR}({\bf v}', {\bf v}) +
\theta' D_{\theta} \theta'/2 +
\theta' \left\{ -\nabla_{t} -
{\cal A}Z_{\cal A} (\theta_{k}\partial_{k}) v_{i}
+ \kappa Z_{\kappa}\partial^{2} \right\} \theta,
\nonumber \\
{\cal S}_{vR}({\bf v}', {\bf v}) =
v 'D_{v}v'/2+ v'\left\{-\nabla_t + \nu Z_{\nu} \partial^{2} \right\}v,
\label{actionR}
\end{eqnarray}
where $D_{v}$ is expressed in renormalized parameters using (\ref{mult}).
It differs from the original (unrenormalized) action (\ref{action}),
(\ref{actionV}) only by the choice of parameters,
${\cal S}_{R} (\Phi,e,\mu) = {\cal S}_{0} (\Phi,e_{0})$, where $e_{0}$ is
the full set of bare parameters and $e$ are their renormalized counterparts.
Thus the original $G=\langle \Phi \dots \Phi \rangle$ and the renormalized
$G_{R}$ Green functions are also related as
$G(e_{0},\dots) = G_{R}(e,\mu,\dots)$; the ellipsis stands for the other
arguments (times/frequencies and coordinates/momenta). We use
$\widetilde{\cal D}_{\mu}$ to denote the differential operation
$\mu\partial_{\mu}$ at fixed bare parameters $e_{0}$ and operate on both
sides of the last relation with it. This gives the basic RG differential
equation:
\begin{eqnarray}
\left\{ {\cal D}_{\mu} - \gamma_{\nu} {\cal D}_{\nu} +
\beta_{g}\partial_{g}+ \beta_{u}\partial_{u}
+ \beta_{\cal A}\partial_{\cal A} \right\} G_{R}=0.
\label{RG}
\end{eqnarray}
Here $u=\kappa/\nu$ and ${\cal D}_{s} = s\partial_{s}$ for any variable $s$.
The RG functions (the $\beta$ functions and the anomalous dimensions
$\gamma$) are defined as
\begin{equation}
\gamma_{F} = \widetilde{\cal D}_{\mu} \ln Z_{F}
\label{gamma}
\end{equation}
for any quantity $F$ and
\begin{eqnarray}
\beta_{g} &=& \widetilde{\cal D}_{\mu} g = g[-y+3\gamma_{\nu}],
\nonumber \\
\beta_{u} &=& \widetilde{\cal D}_{\mu} u = u[\gamma_{\nu}-\gamma_{\kappa}],
\nonumber \\
\beta_{\cal A} &=& \widetilde{\cal D}_{\mu} {\cal A}
= -{\cal A}\gamma_{\cal A}
\label{beta}
\end{eqnarray}
for completely dimensionless variables (coupling constants). Here
$\widetilde{\cal D}_{\mu}$ is the operation ${\cal D}_{\mu}$ at fixed
bare parameters and the second relations in (\ref{beta}) follow from the
definitions and the relations (\ref{mult}). It remains to note that the
differential operator in (\ref{RG}) is nothing other than
$\widetilde{\cal D}_{\mu}$ expressed in renormalized variables.

The one-loop calculation gives:
\begin{eqnarray}
Z_{\nu}=1 - g \bar S_{d} \, \frac{(d-1)}{4(d+2)} \, \frac{1}{y} \, +O(g^{2}),
\quad Z_{\cal A}=1 +O(g^{2}),
\nonumber \\
Z_{\kappa}= 1 - \frac{ g \bar S_{d}}{u(u+1)^{2}} \, \frac{Q}{2d(d+2)} \,
\frac{1}{y} \, +O(g^{2}),
\label{Z}
\end{eqnarray}
where
\begin{eqnarray}
Q = (u+1)(3 {\cal A}^{2} + {\cal A}d - 2{\cal A} +d^{2} -3) -
2{\cal A}({\cal A}-1),
\label{Q}
\end{eqnarray}
$\bar S_{d}\equiv S_{d}/(2\pi)^{d}$ and $S_d=2\pi^{d/2}/\Gamma(d/2)$ is the
surface area of the unit sphere in $d$-dimensional space. Of course, due to
the passivity of the field $\theta$, the constant $Z_{\nu}$ is the same as
in the model (\ref{actionV}); it does not depend on the parameters $u$ and
${\cal A}$ related to the passive field. It is noteworthy that the
expression (\ref{Q}) for $Q$ simplifies for the aforementioned special
values of ${\cal A}$:
\begin{eqnarray}
Q=(u+1)(d-1)(d+2)\ {\rm for}\ {\cal A}=1,
\nonumber \\
Q=(u+1)(d^{2}-3)\ {\rm for}\ {\cal A}=0.
\label{q}
\end{eqnarray}

From (\ref{Z}) we obtain the following explicit one-loop expressions for
the anomalous dimensions:
\begin{eqnarray}
\gamma_{\nu}= g \bar S_{d}\,  \frac{(d-1)}{4(d+2)} +O(g^{2}),
\quad
\gamma_{\cal A}=O(g^{2}),
\nonumber \\
\gamma_{\kappa}=
\frac{ g \bar S_{d}}{u(u+1)^{2}} \, \frac{Q}{2d(d+2)} \,  +O(g^{2}).
\label{explicit}
\end{eqnarray}

In the rapid-change version of our model \cite{amodel}, $Z_{\cal A}=1$ and
$\gamma_{\cal A}=0$ identically because all nontrivial Feynman diagrams of
the 1-irreducible Green function $\langle\theta'v\theta\rangle$ contain
closed circuits of retarded propagators and therefore vanish. In the present
case, the absence of the $O(g)$ term in $Z_{\cal A}$ and $\gamma_{\cal A}$
is a result of the cancellation of the (nontrivial!) contributions from
the three one-loop diagrams in the 1-irreducible Green function
$\langle\theta'v\theta\rangle$. For the counterterm
$\theta' (v\partial) \theta$ such a cancellation is guaranteed by the
Galilean symmetry (to all orders in $g$; see the discussion above). For
$\theta'(\theta\partial)v$ the cancellation looks accidental and
can be explained by a rather simple form of the one-loop diagrams: the
structures corresponding to the counterterm $\theta' (v\partial) \theta$
cancel each other due to the Galilean symmetry, while the structures
corresponding to $\theta'(\theta\partial)v$ enter all the one-loop
diagrams with the same coefficients and cancel out into the
bargain. This mechanism is not expected to work beyound the one-loop
approximation; thus nontrivial contributions of the order $g^{2}$
and higher in $Z_{\cal A}$ and $\gamma_{\cal A}$ are not forbidden.

\section{Fixed points} \label{sec:FP}

It is well known that IR asymptotic behaviour of a multiplicatively
renormalizable field theory is governed by IR attractive fixed points of the
corresponding RG equations. Their coordinates are found from the requirement
that all the $\beta$ functions vanish, $\beta_{i} (g_{*}) =0$,
while the type of the point is determined by the matrix
$\Omega = \{ \Omega_{ik} = \partial\beta_{i}/\partial g_{k} |_{g=g_*}\}$:
for an IR attractive fixed points it is positive, that is, the real parts of
all its eigenvalues are positive. Here $g=\{g_{i}\}$ is the full set of
couplings and $\beta_{i}$ is the full set of the corresponding
$\beta$ functions.

From the explicit expressions (\ref{beta}) and (\ref{explicit}) for
$\beta_{g}$
it follows that the model (\ref{actionV}) has a nontrivial fixed point
\begin{eqnarray}
g_{*} \bar S_{d} = y\, \frac{4(d+2)}{3(d-1)} + O(y^{2}),
\label{fix1}
\end{eqnarray}
which is positive and IR attractive ($\partial_{g}\beta_{g}>0$) for $y>0$
(of course, this fact is well known, see e.g. \cite{Book3,turbo}).
Substituting (\ref{fix1}) into the equation $\beta_{u}=0$ and using the
explicit expressions (\ref{beta}) and (\ref{explicit}) gives
\begin{eqnarray}
2Q  =  u(u+1)^{2} d(d-1),
\label{fix2}
\end{eqnarray}
with $Q$ from (\ref{Q}) and corrections of order $O(y)$.

The last equation is $\beta_{\cal A}=0$. From eqns. (\ref{beta})
and (\ref{explicit}) one finds that it is satisfied
automatically up to the order $O(g)$. Thus there are two possibilities
that cannot be distinguished within the one-loop approximation:

The first one is that $Z_{\cal A}=1$ to all orders in $g$, as it happens
in the rapid-change version of our model \cite{amodel}. Then the equation
$\beta_{\cal A}=0$ becomes an identity and imposes no restriction on the
coordinates of the fixed points. Then eq. (\ref{fix2}) determines the
coordinate $u_{*}$ as a function of the remaining free parameter ${\cal A}$.

In particular, for the most interesting physical case $d=3$, the simple
numerical analysis shows that the positive solution $u_{*}$ of
eq.~(\ref{fix2}) is unique and exists for all ${\cal A}$. As a function
of ${\cal A}$, it achieves a minimum $u_{*}\simeq 0.94$ for
${\cal A}\simeq -0.5$ and grows as $u_{*} = |{\cal A}|+O(1)$ for
${\cal A}\to\pm\infty$. Some special values are $u_{*}\simeq 1.393$ for
${\cal A}=1$ in agreement with the kinematic fixed point of the full-scale
MHD problem \cite{Mns}, $u_{*} =1$ for ${\cal A}=0$ in agreement with
\cite{rebumat} and $u_{*} =1$ for ${\cal A}=-1$. The simple inspection
shows that this fixed point is IR attractive: $\partial_{u}\beta_{u}>0$,
$\partial_{\cal A}\beta_{\cal A}=\partial_{u}\beta_{\cal A}=0$.

A very similar behaviour of the solution $u_{*}({\cal A})$ takes place for
all $d>2$. As a function of $d$, it decreases monotonically and tends to
unity as $d$ tends to infinity. The explicit analytic solution of the cubic
equation (\ref{fix2}) for general $d$ looks rather cumbersome and we do not
present it here. For $d\le2$ our results become inapplicable because the
renormalization of the NS model (\ref{actionV}) itself must be revisited
\cite{turbo}.

Another possibility is that the function $\beta_{\cal A}$ has a nonvanishing
contribution of order $g^{2}$ or higher. Then the equations $\beta_{u} =
\beta_{\cal A}=0$ determine the possible values of the coordinates $u_{*}$
and ${\cal A}_{*}$. To find all their values, the two-loop calculation of
$Z_{\cal A}$ is needed. However, it is clear without any calculation that
${\cal A}_{*}=0$ and ${\cal A}_{*}=1$ are among the possible fixed-point
values of ${\cal A}$ to all orders in $g$: the first case possesses
additional symmetry with respect to the shift $\theta\to\theta+{\rm const}$
(only derivatives of $\theta$ enter the stochastic equation (\ref{1})),
while for the second case the nonlinearity
$V_{i}=(v_{k}\partial_{k})\theta_{i} - (\theta_{k}\partial_{k}) v_{i} =
\partial_{k}(v_{k}\theta_{i}-\theta_{k}v_{i})$ in (\ref{1}) is transverse:
$\partial_{i} V_{i}=0$, so that the nonlocal pressure term (\ref{Pois})
vanishes. The both properties are preserved by the renormalization procedure.

Existence of the fixed points different from ${\cal A}_{*}=0$ and $1$ and
their stability cannot be established without the explicit two-loop
calculation. This issue lies beyound the scope of the present paper; here
we only can say that for the passive vector field advected by the
compressible Kraichnan's ensemble such points do exist; see \cite{TWP}.

\section{Inertial-range anomalous scaling of the correlation functions,
composite fields and operator product expansion} \label{sec:OPE}

The key role in the following is played by the critical dimensions
$\Delta_{n,l}$ associated with the irreducible tensor composite fields
(``local composite operators'' in the field theoretic terminology)
built solely of the fields $\theta$ at a single space-time point
$x=(t,{\bf x})$. They have the forms
\begin{equation}
F_{n,l}\equiv \theta_{i_{1}}(x)\cdots \theta_{i_{l}}(x)\,
\left(\theta_{i}(x)\theta_{i}(x)\right)^{p} + \dots,
\label{Fnl}
\end{equation}
where $l\le n$ is the number of the free vector indices and $n=l+2p$ is
the total number of the fields $\theta$ entering into the
operator; the tensor indices and the argument $x$ of the symbol
$F_{n,l}$ are omitted. The ellipsis stands for the appropriate
subtractions involving the Kronecker delta symbols, which ensure
that the resulting expressions are traceless with respect to
contraction of any given pair of indices, for example,
$\theta_{i}\theta_{j} - \delta_{ij}(\theta_{k}\theta_{k}/d)$ and so on.

The quantities of interest are, in particular, the equal-time pair
correlation functions of the operators (\ref{Fnl}). For these, solving
the corresponding RG equations gives the following asymptotic expression
\begin{equation}
\langle  F_{n,l}(t,{\bf x})F_{k,j}(t,{\bf x}') \rangle \simeq
r^{-\Delta_{n,l}-\Delta_{k,j}}\, \zeta_{n,l;k,j}(mr)
\label{struc}
\end{equation}
with $r=| {\bf x} - {\bf x}'|$ and certain scaling functions $\zeta(mr)$.
To simplify the notation, here and below in similar expressions we
omit the tensor indices and the labels of the scaling functions;
the IR irrelevant parameters (like $\Lambda$ or $\nu_0$) are also not shown.

The last expression in (\ref{struc}) is valid for $\Lambda r \gg 1$
and arbitrary values of $mr$. The inertial-convective range
corresponds to the additional condition that $mr\ll 1$. The forms
of the functions $\zeta(mr)$ are not determined by the RG equations
themselves; their behavior for $mr\to0$ is studied using Wilson's OPE.

According to the OPE, the equal-time product $F_{1}(x)F_{2}(x')$
of two renormalized composite operators at ${\bf x} = ({\bf x}
+ {\bf x'} )/2 = {\rm const}$ and ${\bf r} = {\bf x} - {\bf
x'}\to 0$ can be represented in the form
\begin{equation}
F_{1}(x)F_{2}(x') \simeq \sum_{F} C_{F} ({\bf r}) F(t,{\bf x}),
\label{OPE}
\end{equation}
where the functions $C_{F}$ are the Wilson coefficients, regular
in $m^{2}$, and $F$ are, in general, all possible renormalized local
composite operators allowed by symmetry. More precisely, the
operators entering the OPE are those which appear in the
corresponding Taylor expansions, and also all possible operators
that admix to them in renormalization. If these operators have
additional vector indices, they are contracted with the
corresponding indices of the coefficients $C_{F}$.

It can always be assumed that the expansion in Eq. (\ref{OPE}) is made in
operators with definite critical dimensions $\Delta_{F}$. The correlation
functions (\ref{struc}) are obtained by averaging equation of the type
(\ref{OPE}) with the weight $\exp {\cal S}$, where ${\cal S}$ is the
action functional (\ref{actionR}); the quantities $\langle F \rangle$
appear on the right hand sides. Their asymptotic behavior
for $m\to0$ is found from the corresponding RG equations and
has the form $\langle F \rangle \propto  m^{\Delta_{F}}$.

From the expansion (\ref{OPE}) we therefore
find the following asymptotic expression for the scaling function
$\zeta(mr)$ in the representation (\ref{struc}) for $mr\ll1$:
\begin{equation}
\zeta(mr) \simeq \sum_{F} A_{F}\, (mr)^{\Delta_{F}},
\label{OR}
\end{equation}
where the coefficients $A_{F}=A_{F}(mr)$ are regular in $(mr)^{2}$.

\section{Anomalous scaling and the exponents in the one-loop approximation}
\label{sec:V}

The feature specific of the models of turbulence is the existence of
composite operators with {\it negative} critical dimensions. Such
operators are termed ``dangerous,'' because their contributions to the
OPE diverge at $mr\to0$ \cite{Book3,turbo}.

Obviously, most dangerous are the operators (\ref{Fnl}) with the critical
dimensions $\Delta_{n,l}=-n+O(y)$. Like in the Kraichnan's case, the
analysis shows that their anomalous dimensions can be calculated in the
simplified model without the random forcing in the stochastic equation
(\ref{1}) because the correlator (\ref{2}) does not enter the relevant
Feynman diagrams \cite{MHD1}. Then those operators become multiplicatively
renormalizable, $F_{n,l} = Z_{n,l} F_{n,l}^{R}$. The practical one-loop
calculation of the renormalization constants $Z_{n,l}$ is similar to the
case of Kraichnan's velocity field, discussed in \cite{MHD1,TWP} in detail,
so that here we give only the result:
\begin{eqnarray}
Z_{n,l}= 1 - \frac {g \bar S_{d}}{u(u+1)}\, \frac{{\cal A}^{2} Q_{nl}}
{4d(d+2)} \, \frac{1}{y} +O(g^{2}),
\label{Znl}
\end{eqnarray}
where
\begin{eqnarray}
Q_{n,l} &=& 2n(n-1) -(d+1)(n-l)(d+n+l-2)=
\nonumber \\
&=& - (d-1)(n-l)(d+n+l) + 2l(l-1)
\label{Qnl}
\end{eqnarray}
and $\bar S_{d}$ is defined below equation (\ref{Q}). Note that the same
polynomial $Q_{n,l}$  arises in the scalar case \cite{Q} and in Kraichnan's
MHD model \cite{MHD1}.

The corresponding anomalous dimension is
\begin{eqnarray}
\gamma_{n,l}= \frac {g \bar S_{d}}{u(u+1)}\,
\frac{{\cal A}^{2}Q_{nl}}{4d(d+2)} \, +O(g^{2}).
\label{Gnl}
\end{eqnarray}
Substituting the fixed-point value (\ref{fix1}) gives
\begin{eqnarray}
\gamma_{n,l}=  \frac{{\cal A}^{2}}{u(u+1)} \, \frac{Q_{n,l}}{3d(d-1)} \,
y +O(y^{2}).
\label{Gast}
\end{eqnarray}

If the function $\beta_{\cal A}$ vanishes identically, the solution
$u_{*}({\cal A})$ of the equation (\ref{fix2}) should be substituted into
(\ref{Gast}); then the dependence on the free parameter ${\cal A}$ persists
in $\gamma_{n,l}$. Otherwise, the fixed-point values $u_{*}$, ${\cal A}_{*}$
should be used; see discussion in section~\ref{sec:FP}. In particular, for
${\cal A}_{*}=1$ the fixed-point value of $u_{*}$ is the positive solution
of the quadratic equation $u(u+1)=2(d+2)/d$. Then eq. (\ref{Gast}) becomes
\begin{eqnarray}
\gamma_{n,l}=   \frac{Q_{n,l}}{6(d-1)(d+2)} \, y +O(y^{2}),
\label{Gas}
\end{eqnarray}
which agrees with the result derived in \cite{JJ} for the MHD case. Note
that (\ref{Gas}) coincides with its analog in the Kraichnan's case
\cite{MHD1} up to the substitution $\xi\to y/3$.

For ${\cal A}=0$, the anomalous dimensions $\gamma_{n,l}$ vanish
to all orders in $y$, because the operators $F_{n,l}$ become UV finite and
are not renormalized. In that case, interesting quantities are structure
functions (rather than plain correlation functions); their inertial-range
behaviour is determined by the operators built of the derivatives of the
fields $\theta$; see \cite{matrix,rebumat}.

With this exception, the amplitude ${\cal A}^{2}/u(u+1)$ in (\ref{Gast})
is positive for any physical fixed point. Thus the dimension $\gamma_{n,l}$
is negative for the most interesting case of the scalar operator with
$l=0$ and increases monotonically with $l$ (for a fixed $n$).

From the relation $\Delta_{n,l}=-n+O(y)$ it follows that the critical
dimensions satisfy the same hierarchy relations:
$\Delta_{n,l}>\Delta_{n,l'}$ if $l>l'$, which are conveniently
expressed as inequalities $\partial \Delta_{n,l} / \partial l >0$.

This fact, first established in \cite{Lanotte} for the Kraichnan's MHD model,
has a deep physical meaning: in the presence of large-scale anisotropy, the
leading contribution in the inertial-range behavior $mr\to0$ of the
correlation function like (\ref{struc}) is given by the isotropic ``shell''
($l=0$). The corresponding anomalous exponent is the same as for the purely
isotropic case. The anisotropic contributions give only corrections which
vanish for $mr\to0$, the faster the higher the degree of anisotropy $l$ is.
This effect gives some quantitative support for Kolmogorov's hypothesis of
the local isotropy restoration and appears rather robust, being observed for
the real fluid turbulence \cite{Hier} and the passive scalar model \cite{Q}.

\section{Conclusion} \label{sec:Conc}

We have studied a model of a divergence-free (transverse) vector quantity
$\theta$, passively advected by a random non-Gaussian velocity field with
finite correlation time, governed by the stochastic NS equation.
The model is described by an advection-diffusion equation with
a random large-scale stirring force, nonlocal pressure term
and the most general form of the inertial nonlinearity,
``controlled'' by the parameter ${\cal A}\propto {\cal A}_{0}$.

We have shown that, in the inertial range of scales,
the correlation functions of the field $\theta$ exhibit anomalous scaling
behaviour. The corresponding anomalous exponents are determined
by the critical dimensions of tensor composite fields (\ref{Fnl}) built
solely of the passive vector field. They are calculated (including the
anisotropic sectors) to the leading order of the expansion in $y$, the
exponent entering the correlation function of the stirring force in the
NS equation (the one-loop approximation in the RG terminology).


Like in the special MHD case ${\cal A}={\cal A}_{0}=1$, the exponents
exhibit a kind of hierarchy related to the
degree of anisotropy: the less is the rank of the tensor operator, the
less is the dimension and, consequently, the more important is the
contribution to the inertial-range behaviour. Thus in the presence of
large-scale anisotropy the leading terms, determined by the scalar
operators, remain the same as in the purely isotropic case, in agreement
with the phenomenological hypothesis of the local isotropy restoration.

The question that remains open is whether the amplitude ${\cal A}$
in front of the  ``stretching term'' $(\theta\partial) v$ in the
advection-diffusion equation tends to some fixed-point
values, or it remains a free parameter which the anomalous exponents
depend upon. The analysis of that alternative lies beyound the scope of
the one-loop approximation. We plan to perform it in the nearest future.

\section*{Acknowledgments}

The authors are indebted to Loran Adzhemyan, Michal Hnatich and
Juha Honkonen for discussion.
The authors thank the Organizers of the IV International Conference
``Models in Quantum Field Theory'' dedicated to  A.\,N.\,Vasiliev
(St.~Petersburg--Petrodvorez, 24--27 September 2012)
for the possibility to present the results of this work.
The work was supported in part by the Russian Foundation for Fundamental
Research (project~12-02-00874-a).

\end{document}